# Convergence analysis of the thermal discrete dipole approximation

Sheila Edalatpour[*†], Martin Čuma[**], Tyler Trueax[*], Roger Backman[*] and Mathieu Francoeur[*†]

[*]*Radiative Energy Transfer Lab, Department of Mechanical Engineering, University of Utah, Salt Lake City, UT 84112, USA*

[**]*Center for High Performance Computing, University of Utah, Salt Lake City, UT 84112, USA*

## ABSTRACT

The thermal discrete dipole approximation (T-DDA) is a numerical approach for modeling near-field radiative heat transfer in complex three-dimensional geometries. In this work, the convergence of the T-DDA is investigated by comparison against the exact results for two spheres separated by a vacuum gap. The error associated with the T-DDA is reported for various sphere sizes, refractive indices and vacuum gap thicknesses. The results reveal that for a fixed number of subvolumes, the accuracy of the T-DDA degrades as the refractive index and the sphere diameter to gap ratio increase. A converging trend is observed as the number of subvolumes increases. The large computational requirements associated with increasing the number of subvolumes, and the shape error induced by large sphere diameter to gap ratios, are mitigated by using a nonuniform discretization scheme. Nonuniform discretization is shown to significantly accelerate the convergence of the T-DDA, and is thus recommended for near-field thermal radiation simulations. Errors less than 5% are obtained in 74% of the cases studied by using up to 82712 subvolumes. Additionally, the convergence analysis demonstrates that the T-

[†] Corresponding authors. Tel.: +1 801 581 5721, Fax: +1 801 585 9825

E-mail addresses: mfrancoeur@mech.utah.edu (M. Francoeur), sheila.edalatpour@utah.edu (S. Edalatpour)

DDA is very accurate when dealing with surface polariton resonant modes dominating radiative heat transfer in the near field.

## I. INTRODUCTION

Near-field thermal radiation has recently attracted significant interest due to potential applications in thermal management of nanoscale devices [1], imaging [2], nanomanufacturing [3,4], thermal rectification [5,6], near-field thermal spectroscopy [7-9] and thermophotovoltaic power generation [10-12]. In the near-field regime, arising when the distance between bodies is smaller than Wien's wavelength, radiative heat transfer exceeds the far-field blackbody limit due to tunneling of evanescent modes [13-15]. As such, the classical theory based on Planck's blackbody distribution cannot be applied to near-field thermal radiation predictions. Instead, near-field radiative heat transfer problems are modeled using fluctuational electrodynamics where stochastic current densities representing thermal radiation sources are added to the Maxwell equations [16]. A significant amount of research has been devoted to the analysis of near-field radiative heat transfer in one-dimensional layered geometry for which an exact solution can be derived using dyadic Green's functions (DGFs) [17-21]. Exact solutions have also been derived for configurations such as near-field radiative heat transfer between nanoparticles [22], between two spheres [23-25], and between a sphere and a surface [26,27]. When dealing with three-dimensional complex geometries, it is necessary to employ numerical techniques. So far, a few numerical methods have been proposed for solving the thermal stochastic Maxwell equations [28-34]. Edalatpour and Francoeur [35] presented a relatively simple approach called the thermal discrete dipole approximation (T-DDA). The T-DDA is based on the discrete dipole approximation (DDA), which is a well-known method for modeling light absorption and scattering by particles with size comparable to, or smaller than, the wavelength [36-38]. In both the T-DDA and the DDA, objects are discretized into cubical subvolumes conceptualized as electric point dipoles. The main distinctive feature of the T-DDA is that the dipole moments in the subvolumes are induced not only by an external illumination but also by thermally fluctuating dipoles arising from thermal agitation of charges.

The accuracy and convergence of the DDA have been studied extensively in the literature, and a detailed, comprehensive discussion of this topic can be found in Ref. [38]. The accuracy of the DDA is a function of three main parameters, namely the shape, the size and the refractive index of the objects [38,39]. The convergence of the DDA has been empirically analyzed using analytical solutions for a single, isolated sphere (Mie theory) [37,39-41] and for two spheres in contact [37,42,43]. In general, the accuracy of the DDA degrades as the refractive index and/or the size increase [37,39,40,44], while it improves as the number of subvolumes increases [37,39,44-46]. The computational requirements associated with the DDA grow as the number of subvolumes increases, such that the maximum size and refractive index that can be modeled are limited by the available computational resources. Traditionally, the DDA is said to be suitable for objects with refractive index $m$ satisfying the relation $|m-1| \leq 2$ [47,48]. Larger refractive indices can also be handled with the DDA by utilizing techniques such as the weighted discretization approach [49,50] and the filtered coupled-dipole method [50-52]. Draine and Flatau [47] recommend using the DDA for objects of normalized size $k_0 D_{eff}$ less than 50, where $k_0$ is the magnitude of the wavevector in free space and $D_{eff}$ is the effective diameter of the object. However, this is an approximate criterion since the DDA has been applied to objects as large as $k_0 D_{eff} = 640$ for near unity refractive indices [48]. Additionally, the convergence of the DDA is much faster for cubically-shaped objects that can be represented exactly by cubical subvolumes than for spheres due to the absence of shape error [44,45,53]. Based on an empirical analysis, Draine [39] proposed three criteria for determining the minimum number of subvolumes required to achieve a desired fractional error. These criteria are concerned with the shape error, the variation of the electric field inside the objects, and the minimum subvolume size to ensure that the contribution of the magnetic dipoles to the absorption is negligible when compared to the contribution from the electric dipoles. Zubko et al. [54] pointed out that the second criterion ensuring negligible variation of the electric field overestimates the number of subvolumes required for modeling irregular objects with surface roughness. Yurkin et al. [44,45] analyzed theoretically the convergence of the DDA. It was shown that the error associated with any quantity of interest (e.g., absorption and scattering cross sections) is delimited by a summation of a linear and a quadratic term in the discretization parameter. An extrapolation technique providing an estimation of the error as a function of the discretization parameter was proposed. The superposition of the estimated error and the DDA solution for a cube improved the accuracy of the results by two orders of magnitude. Using this approach, improvement of the accuracy of the DDA was also observed for other shapes.

For thermal radiation problems treated by the T-DDA, the separation gap between the objects is a supplementary parameter that must be accounted for. Indeed, the gap to wavelength ratio and the object size to gap ratio determine the relative contributions of propagating and evanescent modes to radiative heat transfer, and thus the variation of the electric field within the objects. The shape error associated with the T-DDA is also a strong function of the object size to gap ratio. Additionally, Edalatpour and Francoeur [35] showed that Draine's criteria [39] discussed in the previous paragraph largely overestimate the number of subvolumes required to achieve a desired accuracy when applied to the T-DDA. There is therefore a strong motivation for analyzing the accuracy and the convergence of the T-DDA, as near-field thermal radiation simulations differ significantly from traditional scattering and absorption calculations performed with the DDA.

In this work, the convergence of the T-DDA is studied by computing the relative error between the thermal conductance obtained using the exact solution for two spheres separated by a vacuum gap [23-25] and the thermal conductance from T-DDA simulations for the same configuration. The analysis is performed for three types of sizes, namely $k_0D \ll 1$, $k_0D \approx 1$ and $k_0D \gg 1$ ($D$ is the diameter of the sphere). For each size, the distance between the spheres is varied such that the performances of the T-DDA are evaluated in all near-field radiative transfer regimes. As in the DDA, the refractive index of the spheres is expected to have a significant impact on the T-DDA performances. Therefore, various refractive indices, including large and small real and imaginary parts, and a refractive index corresponding to a resonant mode, are examined. A nonuniform discretization scheme is also proposed for accelerating the convergence of the T-DDA.

This paper is organized as follows. The T-DDA described in Ref. [35] has been slightly modified, such that the main steps and equations of the updated formulation are provided in Section II. The approximations made in the T-DDA are listed in Section III. The convergence analysis is afterwards presented and is followed by concluding remarks in Section V. Finally, Appendix A demonstrates that the T-DDA reduces to the previously published dipole approximation [55] in the limit that the sphere diameter is much smaller than the gap size and the wavelength.

## II. DESCRIPTION OF THE T-DDA FORMALISM

The T-DDA framework is established by considering $L$ bodies, with frequency-dependent dielectric functions local in space $\varepsilon_l = \varepsilon_l' + i\varepsilon_l''$ and temperatures $T_l$, submerged in the free space. All bodies are assumed to be in local thermodynamic equilibrium, isotropic and nonmagnetic. $L_e$ bodies emit thermal radiation ($T_l > 0$ K, $l = 1, 2, \ldots, L_e$) while the remaining $L_a$ bodies are pure absorbers ($T_l = 0$ K, $l = L_e + 1, L_e + 2, \ldots, L$). The objective is to calculate the radiative energy transferred to the absorbers. Thermal emission is due to random fluctuations of charges inside the bodies and is modeled using fluctuational electrodynamics [16]. For nonmagnetic materials, as considered here, a fluctuating electric current $\mathbf{J}^{fl}$ is added to Ampère's law in the Maxwell equations. The ensemble average of the fluctuating current (first moment) is zero while the ensemble average of its spatial correlation function (second moment) is given by the fluctuation-dissipation theorem [16]:

$$\langle \mathbf{J}^{fl}(\mathbf{r}',\omega) \otimes \mathbf{J}^{fl}(\mathbf{r}'',\omega') \rangle = \frac{4\omega\varepsilon_0\varepsilon''}{\pi}\Theta(\omega,T)\delta(\mathbf{r}'-\mathbf{r}'')\delta(\omega-\omega')\bar{\bar{\mathbf{I}}} \qquad (1)$$

where $\otimes$ denotes the outer product defined as the multiplication of the first vector by the conjugate transpose of the second vector, $\bar{\bar{\mathbf{I}}}$ is the unit dyadic and $\Theta(\omega,T)$ is the mean energy of an electromagnetic state given by $\hbar\omega/[\exp(\hbar\omega/k_B T) - 1]$. Due to the random nature of the fluctuating current, near-field thermal radiation problems are stochastic and are mathematically described by the thermal stochastic Maxwell equations.

The total electric field at location $\mathbf{r}$ and frequency $\omega$ is the sum of contributions from fluctuating, scattered and incident fields. The fluctuating field is generated by thermal excitation of charges in bodies with temperature larger than absolute zero, while the scattered field is due to multiple electromagnetic interactions between the bodies. The incident field is produced by an external source such as thermal emission by the surroundings (sometimes referred to as the bosonic field or the thermal bath) and/or illumination by a laser. The following free space vector wave equation for the total electric field **E** is derived from the thermal stochastic Maxwell equations:

$$\nabla \times \nabla \times \mathbf{E}(\mathbf{r},\omega) - k_0^2 \mathbf{E}(\mathbf{r},\omega) = i\omega\mu_0 \mathbf{J}(\mathbf{r},\omega) \tag{2}$$

The current **J** is an equivalent source function producing scattered and fluctuating fields:

$$\mathbf{J}(\mathbf{r},\omega) = -i\omega\varepsilon_0(\varepsilon-1)\mathbf{E}(\mathbf{r},\omega) + \mathbf{J}^{fl}(\mathbf{r},\omega) \tag{3}$$

where the first term on the right-hand side of Eq. (3) is the source function for the scattered field [35]. The incident field is generated by an external source and satisfies the homogenous vector wave equation in free space $\nabla \times \nabla \times \mathbf{E}^{inc}(\mathbf{r},\omega) - k_0^2 \mathbf{E}^{inc}(\mathbf{r},\omega) = \mathbf{0}$. The total electric field at location **r** and frequency $\omega$ can thus be written as follows:

$$\mathbf{E}(\mathbf{r},\omega) = i\omega\mu_0 \int_V \overline{\overline{\mathbf{G}}}(\mathbf{r},\mathbf{r}',\omega) \cdot \mathbf{J}(\mathbf{r}',\omega) dV' + \mathbf{E}^{inc}(\mathbf{r},\omega) \tag{4}$$

where $V$ is the total volume of the emitting and absorbing bodies and $\overline{\overline{\mathbf{G}}}$ is the free space DGF defined as [56]:

$$\overline{\overline{\mathbf{G}}}(\mathbf{r},\mathbf{r}',\omega) = \frac{e^{ik_0 R}}{4\pi R}\left[\left(1 - \frac{1}{(k_0 R)^2} + \frac{i}{k_0 R}\right)\overline{\overline{\mathbf{I}}} - \left(1 - \frac{3}{(k_0 R)^2} + \frac{3i}{k_0 R}\right)\hat{\mathbf{R}} \otimes \hat{\mathbf{R}}\right] \tag{5}$$

The first term on the right-hand side of Eq. (4) is the sum of the fluctuating field ($i\omega\mu_0 \int_V \overline{\overline{\mathbf{G}}}(\mathbf{r},\mathbf{r}',\omega) \cdot \mathbf{J}^{fl}(\mathbf{r}',\omega) dV'$), due to thermal fluctuations everywhere in $V$ where $T > 0$ K, and the scattered field ($k_0^2 \int_V \overline{\overline{\mathbf{G}}}(\mathbf{r},\mathbf{r}',\omega) \cdot (\varepsilon-1)\mathbf{E}(\mathbf{r}',\omega)dV'$). In Eq. (5), $R = |\mathbf{r}-\mathbf{r}'|$ and $\hat{\mathbf{R}} = (\mathbf{r}-\mathbf{r}')/|\mathbf{r}-\mathbf{r}'|$.

Equation (4) is discretized by dividing the $L$ objects into $N$ cubical subvolumes on a cubical lattice. The electric field at the center point $\mathbf{r}_i$ of subvolume $i$ ($i = 1, 2, \ldots, N$) can be written as:

$$\mathbf{E}(\mathbf{r}_i,\omega) = i\omega\mu_0 \sum_{j \neq i} \int_{\Delta V_j} \overline{\overline{\mathbf{G}}}(\mathbf{r}_i,\mathbf{r}',\omega) \cdot \mathbf{J}(\mathbf{r}',\omega)dV' + i\omega\mu_0 \int_{\Delta V_i} \overline{\overline{\mathbf{G}}}(\mathbf{r}_i,\mathbf{r}',\omega) \cdot \mathbf{J}(\mathbf{r}',\omega)dV' + \mathbf{E}^{inc}(\mathbf{r}_i,\omega) \qquad (6)$$

The integration over $\Delta V_i$ in Eq. (6) is treated separately since the DGF has a singularity at $\mathbf{r}' = \mathbf{r}_i$. Next, it is assumed that the free space DGF and the electric field are constant within each subvolume. The only exception arises for the integration of the DGF over $\Delta V_i$, where the principal value method is used [57]. Note that the validity of these approximations is discussed in Section III. Equation (6) then becomes:

$$\mathbf{E}_i = i\omega\mu_0 \sum_{j \neq i} \overline{\overline{\mathbf{G}}}_{ij} \cdot \int_{\Delta V_j} \mathbf{J}(\mathbf{r}',\omega)dV' + \frac{i}{3\omega\varepsilon_0}\left[2\left(e^{ik_0 a_i}(1-ik_0 a_i)-1\right)-1\right]\left[\frac{1}{\Delta V_i}\int_{\Delta V_i} \mathbf{J}(\mathbf{r}',\omega)dV'\right] + \mathbf{E}_i^{inc} \qquad (7)$$

where $\overline{\overline{\mathbf{G}}}_{ij}$ is the free space DGF between $\mathbf{r}_i$ and $\mathbf{r}_j$, and $a_i$ (= $(3\Delta V_i/4\pi)^{1/3}$) is the effective radius of subvolume $i$. When evaluating the integration over $\Delta V_i$ in Eq. (6), two assumptions are made. First, the current $\mathbf{J}$ inside subvolume $i$ is approximated by its volumetric average. Secondly, when applying the principal value method, subvolume $i$ is approximated as a sphere of equivalent volume. The validity of this approximation has been verified by comparison against an exact method [58]. A more rigorous approach for treating the singularity of the DGF can be found in Refs. [58,59].

Under the assumption that the subvolumes are small compared to the wavelength, it is reasonable to model a subvolume as an electric point dipole. A given subvolume $i$ is therefore characterized by a total dipole moment $\mathbf{p}_i$ that is related to the equivalent current via the relation $\mathbf{p}_i = i \int_{\Delta V_i} \mathbf{J}(\mathbf{r}',\omega) dV'/\omega$ [60], where $i$ is the complex constant. After substitution of Eq. (3), the following expression for the total dipole moment is determined:

$$\mathbf{p}_i = \Delta V_i \varepsilon_0 (\varepsilon_i - 1) \mathbf{E}_i + \frac{i}{\omega} \int_{\Delta V_i} \mathbf{J}^{fl}(\mathbf{r}',\omega) dV' \tag{8}$$

The first term on the right-hand side of Eq. (8) is the induced dipole moment $\mathbf{p}_i^{ind}$ while the second term is the thermally fluctuating dipole moment $\mathbf{p}_i^{fl}$. Since the fluctuating current is converted into a thermally fluctuating dipole moment, it is more appropriate to express the fluctuation-dissipation theorem as follows:

$$\left\langle \mathbf{p}_i^{fl}(\omega) \otimes \mathbf{p}_i^{fl}(\omega') \right\rangle = \frac{4 \varepsilon_0 \varepsilon_i'' \Delta V_i}{\pi \omega} \Theta(\omega, T) \delta(\omega - \omega') \overline{\overline{\mathbf{I}}} \tag{9}$$

The main equation of the T-DDA is derived by writing Eq. (7) in terms of dipole moments:

$$\frac{1}{\alpha_i} \mathbf{p}_i - \frac{k_0^2}{\varepsilon_0} \sum_{j \neq i} \overline{\overline{\mathbf{G}}}_{ij} \cdot \mathbf{p}_j = \frac{3}{(\varepsilon_i + 2)} \frac{1}{\alpha_i^{CM}} \mathbf{p}_i^{fl} + \mathbf{E}_i^{inc} \tag{10}$$

The variables $\alpha_i^{CM}$ and $\alpha_i$ are the Clausius–Mossotti and radiative polarizabilities of dipole $i$ given by:

$$\alpha_i^{CM} = 3 \varepsilon_0 \Delta V_i \frac{\varepsilon_i - 1}{\varepsilon_i + 2} \tag{11a}$$

$$\alpha_i = \frac{\alpha_i^{CM}}{1-(\alpha_i^{CM}/2\pi\varepsilon_0 a_i^3)[e^{ik_0 a_i}(1-ik_0 a_i)-1]} \quad (11b)$$

Note that the fluctuation-dissipation theorem for the fluctuating dipole moment given by Eq. (9) is different from the expression previously reported in the literature [55]. This is explained by the fact that the induced dipole moment due to the interaction of subvolume $i$ with itself is implicitly included in the fluctuation-dissipation theorem of Ref. [55]. In the current formulation, the induced dipole moment due to self-interaction of subvolume $i$ is accounted for in the first term on the left-hand side of Eq. (10). It is shown in Appendix A that the fluctuation-dissipation theorem given by Eq. (9) combined with Eq. (10) is equivalent to the formulation presented in Ref. [55].

Equation (10) is a system of $3N$ scalar equations that can be written in a compact matrix form as follows:

$$\overline{\overline{\mathbf{A}}}\cdot\overline{\mathbf{P}} = \overline{\mathbf{E}}^{fdt} + \overline{\mathbf{E}}^{inc} \quad (12)$$

where $\overline{\overline{\mathbf{A}}}$ is the $3N$ by $3N$ deterministic interaction matrix [35], $\overline{\mathbf{E}}^{fdt}$ is a $3N$ stochastic column vector containing the first term on the right-hand side of Eq. (10) and its correlation matrix is obtained using the fluctuation-dissipation theorem, $\overline{\mathbf{E}}^{inc}$ is a $3N$ deterministic column vector containing the incident field and $\overline{\mathbf{P}}$ is a $3N$ stochastic column vector containing the unknown total dipole moments.

The monochromatic power dissipated in the absorbers is given by [37,61]:

$$\langle Q_{abs,\omega}\rangle = \frac{\omega}{2}\sum_{i\in abs}\left(\mathrm{Im}[(\alpha_i^{-1})^*]-\frac{2}{3}k_0^3\right)\mathrm{tr}\left(\langle \mathbf{p}_i^{ind}\otimes\mathbf{p}_i^{ind}\rangle\right) \quad (13)$$

where the superscript * denotes complex conjugate while $\text{tr}\langle \mathbf{p}_i^{ind} \otimes \mathbf{p}_i^{ind} \rangle$ is the trace of the auto-correlation function of the induced dipole moment of subvolume *i*. The summation in Eq. (13) is performed strictly over the subvolumes contained within the absorbers. Since the absorbers are at a temperature of 0 K, $\mathbf{p}_i^{fl} = \mathbf{0}$ and $\langle \mathbf{p}_i \otimes \mathbf{p}_i \rangle = \langle \mathbf{p}_i^{ind} \otimes \mathbf{p}_i^{ind} \rangle$ such that Eq. (13) can be calculated directly from the system of equations (12). The trace of the auto-correlation function of the total dipole moment is determined using the correlation matrix of $\overline{\mathbf{P}}$ obtained from Eq. (12) [35,62]:

$$\langle \overline{\mathbf{P}} \otimes \overline{\mathbf{P}} \rangle = \overline{\overline{\mathbf{A}}}^{-1} \cdot \langle (\overline{\mathbf{E}}^{fdt} \otimes \overline{\mathbf{E}}^{fdt}) + (\overline{\mathbf{E}}^{inc} \otimes \overline{\mathbf{E}}^{inc}) \rangle \cdot \left( \overline{\overline{\mathbf{A}}}^{-1} \right)^{\dagger} \qquad (14)$$

where the superscript † is the Hermitian operator. The fact that $\overline{\mathbf{E}}^{fdt}$ and $\overline{\mathbf{E}}^{inc}$ are uncorrelated (i.e., $\langle \overline{\mathbf{E}}^{fdt} \otimes \overline{\mathbf{E}}^{inc} \rangle = \overline{\overline{\mathbf{0}}}$) has been used when deriving Eq. (14). The correlation matrix $\langle \overline{\mathbf{E}}^{fdt} \otimes \overline{\mathbf{E}}^{fdt} \rangle$ is derived by applying the fluctuation-dissipation theorem given by Eq. (9).

The difference between the T-DDA framework described here and in Ref. [35] comes from the splitting of the fluctuating field and the incident field. In this paper, the incident field represents solely the field produced by external sources. Additionally, the system of equations (12) is written in terms of total dipole moments rather than in terms of induced dipole moments.

As a final remark, it is important to recognize that the system of equations (12) is stochastic and can be solved in different ways. Hereafter, the computations are performed in a deterministic manner by calculating directly the dipole auto-correlation function from Eq. (14). Alternatively, Eq. (12) can be solved directly by assuming that only one subvolume is thermally emitting while all other subvolumes are at a temperature of 0 K. These calculations need to be repeated for each subvolume contained in the emitters, and the dipole auto-correlation function can thus be determined. The correlation matrix method is attractive as it does not involve multiple solutions of a system of equations. On the other hand, this methodology is computationally expensive due to large memory requirements when dealing with a large number of subvolumes. More details on this topic will be provided in Section IV.

### III. APPROXIMATIONS ASSOCIATED WITH THE T-DDA

Following the derivation presented in Section II, the approximations made in the T-DDA can be summarized into four points.

1. *Discretization of the objects into cubical subvolumes*. The error introduced by this approximation is called the shape error [39,44]. The shape error is nonexistent for objects that can be represented exactly by a cubical lattice such as a cube [44], while it can be large for curved objects such as a sphere. The shape error for multiple objects closely spaced from each other, or in contact, is larger than for a single object. As discussed later, this is related to the importance of representing accurately the gap size between discretized objects. The extent to which the shape error negatively affects the accuracy of the results is a strong function of the refractive index of the object [39]. A large refractive index implies a high contrast between the object and the free space, which amplifies the shape error. Approximating objects by a cubical lattice is valid when the size of the subvolumes is small compared to the characteristic lengths of the problem, namely the size of the objects and their separation distance.

2. *Constant electric field in each subvolume*. Radiative heat transfer in the near field occurs via propagating and evanescent modes. When dealing with propagating modes, approximating the electric field as constant within a subvolume is acceptable when the size of the subvolume is smaller than the free space wavelength ($\lambda$), smaller than the material wavelength ($\lambda_m = \lambda/\text{Re}(m)$), and smaller than the decay length of the electric field ($\lambda/\text{Im}(m)$). For evanescent modes, the approximation of constant electric field within a subvolume is acceptable when the size of the subvolume is small compared to the radiation penetration depth. The penetration depth of evanescent modes ranges from $\lambda_m$ to the thickness of the gap separating the objects. Furthermore, for objects with sharp edges such as cubes, the size of the subvolumes must be small compared to the characteristic length of the object, even if the object is much smaller than the wavelength. This ensures that the large electric field gradients near the edges are accurately represented [53].

3. *Constant free space DGF inside each subvolume.* The variations of the free space DGF $\overline{\overline{\mathbf{G}}}(\mathbf{r},\mathbf{r}',\omega)$ inside the objects are proportional to $\lambda/R$, where $R$ is the distance between points $\mathbf{r}$ and $\mathbf{r}'$ [59]. When the size of the object is much smaller than the wavelength (Rayleigh regime), sharp variations of the free space DGF arise inside the object. As such, the validity of this assumption becomes questionable in the Rayleigh regime [58]. In addition, the free space DGF in Eq. (10) is multiplied by the dielectric function $\varepsilon$, such that the accuracy of this approximation degrades with increasing the refractive index $m$ ($m = \sqrt{\varepsilon}$).

4. *Integration of the free space DGF, $\overline{\overline{\mathbf{G}}}(\mathbf{r}_i,\mathbf{r}',\omega)$, over the subvolume $\Delta V_i$ where the singularity of the DGF is located (second term on the right-hand side of Eq. (6)).* To be able to perform the principal volume integral analytically, the cubical subvolume $i$ is approximated by a sphere of equivalent volume [57,63]. The radiative polarizability discussed in section II (Eq. (11b)) is a result of this assumption [35]. Rather than performing the principal volume integration, different polarizability models based on physical arguments have been proposed [39,64-69].

The impact of these approximations on the accuracy of the T-DDA depends on the specific parameters of the problem. This is discussed in the next section.

## IV. ACCURACY OF THE T-DDA

The accuracy of the T-DDA is assessed by comparison against exact results for two spheres [23-25]. As shown in Fig. 1, two spheres of same diameter $D$ and same refractive index $m$ are separated by a distance $X(y,z)$ and are exchanging thermal radiation in the free space. The minimum gap size between the spheres is denoted by $d$ ($= X(0,0)$). For simplicity, it is assumed that there is no incident field. Sphere 1 is at a temperature $T + \delta T$, while sphere 2 is maintained at a temperature $T$. The spectral thermal conductance at temperature $T$ and angular frequency $\omega$ is given by:

$$G_\omega(T) = \lim_{\delta T \to 0} \frac{\langle Q_{net,\omega} \rangle}{\delta T} \tag{15}$$

where $\langle Q_{net,\omega} \rangle = \langle Q_{abs,\omega,12} \rangle - \langle Q_{abs,\omega,21} \rangle$ is the net spectral heat rate. The power dissipated in sphere 2 due to thermal emission by sphere 1, $\langle Q_{abs,\omega,12} \rangle$, is calculated from Eq. (13). Due to reciprocity, the power dissipated in sphere 1 due to thermal emission by sphere 2 can be computed as $\langle Q_{abs,\omega,21} \rangle = \langle Q_{abs,\omega,12} \rangle \frac{\Theta(\omega,T)}{\Theta(\omega,T+\delta T)}$. Therefore, the spectral conductance at temperature $T$ is obtained solely from $\langle Q_{abs,\omega,12} \rangle$:

$$G_\omega(T) = \frac{\langle Q_{abs,\omega,12} \rangle}{\Theta(\omega,T)} \frac{\partial \Theta(\omega,T)}{\partial T} \tag{16}$$

Hereafter, sphere 1 is referred to as the emitter while sphere 2 is called the absorber. The spectral thermal conductance is calculated at a temperature of 300 K and at an angular frequency of $1.884 \times 10^{14}$ rad/s. This corresponds to a vacuum wavelength of 10 μm, which is roughly the dominant wavelength emitted by a body at 300 K.

Approximate solutions for the two-sphere problem have been proposed in the literature for two limiting cases. The proximity approximation is applicable when the size of the spheres is much larger than their separation gap ($D \gg d$) [25,27]. For this case, the conductance between the spheres is calculated as a summation of local heat transfer coefficients between two semi-infinite media separated by different gap sizes. The second limiting case is the dipole approximation, which is valid when the size of the spheres is much smaller than the wavelength while their separation gap is a few times larger than their diameter ($D \ll \lambda$ and $d \gg D$) [55]. In the dipole approximation, the contributions from the quadrupoles and higher order poles as well as multiple scattering between the spheres are neglected. Appendix A demonstrates that the T-DDA reduces to the dipole approximation when $D \ll \lambda$ and $d \gg D$.

The accuracy of the T-DDA is evaluated for three sizes, $k_0 D \ll 1$, $k_0 D \approx 1$ and $k_0 D \gg 1$. For each size, various gap distances in the near-field regime of thermal radiation (i.e., $d < \lambda$) are considered. Overall, a total of seven cases, summarized in Table I, are investigated. For each case listed in Table I, the convergence of the T-DDA is analyzed for six different refractive indices (see Table II), including high and low real and imaginary parts, and a refractive index corresponding to surface phonon-polariton resonance of a silica sphere. The spectral thermal conductance between the spheres is calculated with the T-DDA using various discretization sizes and is compared against exact results. All computations were performed with a hybrid OpenMP-MPI parallel T-DDA Fortran code utilizing the ScaLAPACK library as implemented in the Intel

Math Kernel Library for the interaction matrix inversion. The computational time for the largest number of subvolumes used in case 1 (73824) is approximately 18.7 hours when run on 150 nodes each having 2 six-core Intel Xeon X5660 processors with a speed of 2.80 GHz. This amounts in a total of 33660 service units (i.e., core-hours). Note that 99.8% of the aforementioned time is devoted to the calculation of the inverse of the interaction matrix.

### A. Regime $k_0D \ll 1$

Two sizes, namely $k_0D = 0.00628$ ($D = 10$ nm) and $k_0D = 0.0943$ ($D = 150$ nm), are investigated. A gap thickness of $d/\lambda = 0.001$ ($d = 10$ nm) is selected for $k_0D = 0.00628$ (case 1 in Table I). Larger gaps correspond to the dipolar regime for which a closed-form expression exists (see Appendix A) [55], while the validity of the fluctuational electrodynamics framework is questionable at sub-10 nm gaps. Two gap sizes of $d/\lambda = 0.001$ ($d = 10$ nm) and $d/\lambda = 0.015$ ($d = 150$ nm) are tested for $k_0D = 0.0943$ (cases 2 and 3 in Table I, respectively).

The absolute value of the relative error of the conductance as a function of the number of subvolumes per sphere and the refractive index is provided in Fig. 2 for case 1. The dashed line shows the 5%-error threshold. It can be seen that for all refractive indices, describing a sphere by a single subvolume results in an error of approximately 30% even if $D \ll \lambda$. This is due to the shape error and the non-negligible variation of the electric field inside the spheres. The shape error is caused by an inaccurate representation of the separation distance $X(y,z)$ between the spheres that is fixed at 10 nm, while, in reality, it should vary from 10 nm to 20 nm. Clearly, the subvolume size to gap ratio, $\Delta/d$, should be much smaller than unity in order to minimize the shape error. Additionally, heat transfer in case 1 is dominated by exponentially decaying evanescent modes with minimum penetration depth approximately equal to the gap size $d$. This results in sharp variations of the electric field within the spheres with diameter $D$ equals to $d$. The variation of the electric field is obviously not taken into account when modeling a sphere by a single subvolume. As for the shape error, the ratio $\Delta/d$ must be much smaller than unity in order to represent accurately the variation of the electric field within the spheres. Physically, this error can be understood by recognizing that when $d >> D$ is not satisfied, the multipoles inside the spheres are excited by the evanescent modes [70] such that the dipole approximation is inapplicable. When $d/\lambda$ is increased to 10 ($d$ = 100 µm), which results in a ratio $\Delta/d$ of 0.0000806, describing each sphere by a single subvolume leads to a small error of 0.01% (result not shown). This is to be expected, since the 10 nm variations of the distance $X(y,z)$ between the spheres along the $y$- and $z$-axis are insignificant compared to the gap $d$ of 100 µm. Additionally, the electric field within the spheres is nearly uniform as heat transfer occurs via propagating modes and $D << \lambda$.

In Fig. 2, the error grows as the number of subvolumes is increased from one to eight, and then decreases as the number of subvolumes is further increased. This counterintuitive behavior has also been observed for a large gap size of $d/\lambda = 10$ and for DDA simulations of a single sphere of size $k_0D = 0.00628$ (results not shown). The shape error and the error associated with the assumption of constant electric field within the subvolumes both decrease when increasing the number of subvolumes. However, an additional error caused by the sharp variation of the free space DGF inside the spheres comes into picture. The free space DGF between points $\mathbf{r}_i$ and $\mathbf{r}_j$, $\overline{\overline{\mathbf{G}}}_{ij}$, varies rapidly as $\mathbf{r}_j$ approaches $\mathbf{r}_i$ and becomes singular when $\mathbf{r}_i = \mathbf{r}_j$. Since the spheres are much smaller than the wavelength, the points $\mathbf{r}_i$ and $\mathbf{r}_j$ are always close to each other which results in sharp variations of the DGF throughout the entire spheres. The variations of the free space DGF within small objects do not introduce any error in the T-DDA when a single subvolume per sphere is used, as the integration of the DGF over the subvolume is performed analytically (see Eq. (7)). The variations of the free space DGF induce an error when modeling the objects with more than one subvolume. Note that this error was also observed in Refs. [37,58] when applying the DDA to Rayleigh particles. Chaumet et al. [58] showed that performing the integration of the free space DGF over the subvolumes, instead of assuming constant free space DGF, improves the accuracy of the DDA for very small spherical particles. The assumption of constant free space DGF inside the subvolumes, and therefore the T-DDA results, become more accurate as the number of subvolumes increases.

As expected, the error strongly depends on the refractive index of the material. The error grows as both the real and the imaginary parts of the refractive index increase. In general, increasing the refractive index negatively affects the accuracy of the T-DDA by amplifying the shape error [39], by amplifying the error associated with assuming the free space DGF constant within the subvolumes, and by increasing the variation of the electric field inside the spheres. For case 1, the fact that the error increases with increasing the refractive index is mostly due to the amplification of the shape error and the variation of the DGF; the refractive index has only a small influence on the variation of the electric field within the spheres since this variation is caused by evanescent modes with minimum penetration depth approximately equal to the gap size $d$. The amplification of the shape and constant DGF errors with increasing the refractive index can be mitigated by increasing the number of subvolumes, as shown in Fig. 2. The refractive index $m_f$ corresponds to surface phonon-polariton resonance of a silica sphere. In the near field, the total thermal conductance is largely dominated by the contribution of surface phonon-polaritons [23]. Here, the conductance for $m_f$ is one to six orders of magnitude larger than the conductance calculated for the refractive indices $m_a$ to $m_e$. As depicted in Fig. 2, the T-DDA converges rapidly for the resonant refractive index. Furthermore, the spectral locations of the resonant modes are predicted accurately via the T-DDA [35]. This demonstrates that the T-DDA is an accurate tool for predicting surface phonon-polariton mediated near-field radiative heat transfer.

In case 1, errors of 1.2% and 1.8% are obtained using 17256 subvolumes for $m_a$ and $m_b$, respectively. Reducing the error to less than 5% for other refractive indices requires a larger number of subvolumes. The number of subvolumes used for the simulations is limited by the memory requirement for storing the interaction matrix $\bar{\bar{A}}$. For $N$ subvolumes, $144N^2$ bytes of memory are needed to store $9N^2$ complex elements of the interaction matrix with a double precision format. The size of the subvolumes $\Delta$ decreases with $N$ as $\Delta \propto N^{-1/3}$. This implies that the memory requirement is proportional to $\Delta^{-6}$. Additionally, the computational time, which is almost equal to the calculation time of the inverse of $\bar{\bar{A}}$, is approximately proportional to $N^3$. A significant amount of computational resources is therefore required when a large number of subvolumes is used. A solution to this bottleneck is the implementation of a nonuniform discretization scheme. Indeed, depending on the problem, some portions of the spheres may have negligible contribution to the overall heat transfer. A cross section of the spatial distribution of the normalized volumetric power absorbed by sphere 2 is shown in Fig. 3 for the refractive index $m_c$ and for 39024 uniform subvolumes. The cross section is parallel to the $x$-$y$ plane and passes through the center of the sphere. For this case, more than 85% and 95% of the absorption takes place within distances of 6.18 nm and 7.84 nm from the left edge of the absorbing sphere, respectively. Also, the shape error is more important for the portion of the absorber facing the emitter. Based on the power distribution of Fig. 3, a nonuniform discretization with 36168 subvolumes is proposed in Fig. 4, where the size of the subvolumes increases as the power absorbed decreases. The error obtained with 36168 nonuniform subvolumes per sphere is 4.7%, as opposed to 8.16% when using 39024 uniform subvolumes per sphere. Clearly, a nonuniform discretization scheme helps achieving a smaller error for a fixed number of subvolumes. The distribution of the fine and coarse subvolumes in a nonuniform discretization depends on the physics of the problem. Performing a preliminary simulation using a reasonable amount of uniform subvolumes (e.g., 5000) is helpful for visualizing the power distribution and thus defining an adequate nonuniform meshing. Nonuniform discretization has also been applied to refractive indices $m_d$ to $m_f$. Note that the error values reported in all figures are shown with open symbols when uniform discretization is utilized, while filled symbols are used for nonuniform discretization. Errors less than 5% are achieved for $m_d$ and $m_f$ using nonuniform discretization. The smallest error obtained for $m_e$ is 11.1% with 59360 nonuniform subvolumes. It should be noted that both the real and the imaginary parts of $m_e$ are large. Therefore, the shape error and the variation of the free space DGF are larger for $m_e$ than for the other refractive indices.

The absolute value of the relative error for case 2 is plotted in Fig. 5. Compared to case 1, the size of the spheres is increased while the separation gap $d$ is the same. The error follows the same trend as in case 1. However, for a given number of subvolumes, the errors in Fig. 5 are considerably larger than in case 1 except for the resonant refractive index $m_f$. Increasing the ratio $D/d$ while keeping the number of subvolumes fixed results in a larger discretization size such that the accuracy of all approximations listed in Section III deteriorates. Modeling each sphere by a single subvolume also leads to larger errors when compared to case 1, due to the fact that the shape error and the variation of the electric field within the spheres increase as the ratio $D/d$ increases (for a fixed number of subvolumes, $\Delta/d$ increases when compared to case 1). As for case 1, the thermal conductance at resonance is well predicted by the T-DDA and an error of 2.0% is achieved with 33552 uniform subvolumes per sphere. Nonuniform discretization is adopted for the other refractive indices. An error less than 5% is achieved for the refractive indices $m_a$ to $m_d$ when using up to 33740 nonuniform subvolumes. For the largest refractive index $m_e$, the error reduces from 64.9% with 33552 uniform subvolumes to 10.7% with 59408 nonuniform subvolumes. A better accuracy can be obtained for $m_e$ by increasing further the numbers of subvolumes. For all refractive indices considered in case 2, more than 95% of the absorption occurs within the first half of the sphere. This shows that as the ratio $D/d$ increases, a smaller portion of the absorber contributes to the overall heat exchange, since radiative transfer is dominated by evanescent modes with minimum penetration depth $d$ that is much smaller than the sphere diameter $D$. As such, nonuniform discretization can effectively be utilized in this situation.

Next, the gap thickness is increased to $d/\lambda = 0.015$ while the size is kept constant at $k_0 D = 0.0943$ (case 3). The absolute value of the relative error is shown in Fig. 6. It can be seen that the error obtained in case 3 is very similar to case 1. Indeed, in both cases 1 and 3, radiative transfer is dominated by evanescent modes ($d << \lambda$) and the sphere diameter to gap ratio, $D/d$, is the same. This implies that the shape error and the error associated with the assumption of constant electric field within the subvolumes introduce the same amount of inaccuracy in cases 1 and 3 as the ratios $\Delta/d$ and $\Delta/D$ are the same for a fixed number of subvolumes. Additionally, the error associated with assuming the free space DGF as constant inside the subvolumes is still important since $D << \lambda$. An error less than 5% is obtained for $m_a$ and $m_b$ when 17256 uniform subvolumes are used. For the other refractive indices, nonuniform discretization has been applied. The error reduces to less than 5% for refractive indices $m_c$, $m_d$, and $m_f$ with up to 35256 nonuniform subvolumes. As in the previous cases, the most difficult refractive index to handle is $m_e$ for which an error of 10.3% is achieved using 74180 nonuniform subvolumes.

## B. Regime $k_0 D \approx 1$

The convergence of the T-DDA is analyzed for $k_0 D = 1.01$ ($D = 1.6$ µm) and two gap sizes of $d/\lambda = 0.001$ ($d = 10$ nm) and $d/\lambda = 0.1$ ($d = 1$ µm), corresponding to cases 4 and 5, respectively. The absolute value of the relative error for case 4 is shown in Fig. 7, where up to 33552 uniform subvolumes per sphere are used. For this number of uniform subvolumes and for the refractive indices $m_c$, $m_d$, and $m_e$, the error is extremely large; as such, these points are not plotted in Fig. 7. This behavior can be explained by analyzing the discretized spheres shown in Fig. 8(a). Clearly, the size of the subvolumes is too large for a 10-nm-thick gap ($\Delta/d = 4.0$), such that the shape error is significant. The discretization of the spheres should be fine enough compared to the gap size ($\Delta/d \ll 1$) in order to represent accurately the smooth variation of $X(y,z)$ with respect to the $y$- and $z$-axis. A nonuniform discretization using smaller subvolumes at the front sides of the spheres is thus beneficial. This is particularly helpful because the size of the spheres is much larger than the gap size ($D/d = 160$) such that only a small portion of the absorber contributes to the overall heat transfer (more than 95% of the absorption takes place within a distance smaller than 240 nm for all refractive indices). A nonuniform discretization with 43324 subvolumes is applied to case 4 and is shown in Fig. 8(b), where $\Delta/d = 0.4$ for $d = 10$ nm. The variation of $X(y,0)$ between the discretized spheres along the $y$-axis for the uniform and nonuniform discretizations is shown in Fig. 8(c). The distance $X(y,0)$ varies in a smoother manner with the nonuniform discretization scheme compared to the uniform discretization. Consequently, by decreasing the shape error, the accuracy of the T-DDA is drastically improved. With the nonuniform discretization, an error less than 3.1% is obtained for $m_a$, $m_b$ and $m_f$. Also, compared to the uniform discretization with 33552 subvolumes, the error reduces from 310% to 15.2% for $m_c$, from 332.5% to 17.5% for $m_d$ and from 867.7% to 40.5% for $m_e$. The error for $m_e$ is decreased further to 29.5% using 77196 nonuniform subvolumes. It is worth noting that for the small refractive index $m_a$, an error of 5.4% is achieved with a simple uniform discretization of 33552 subvolumes ($\Delta/d = 4.0$). Therefore, it can be concluded that the value of $\Delta/d$ required for convergence depends strongly on the refractive index of the material. This is in agreement with the observation made in the DDA that the shape error is a function of the refractive index [39].

The absolute value of the relative error for case 5 is presented in Fig. 9. Compared to case 4, the gap size is increased while the sphere size is the same ($D/d$ is 1.6 instead of 160). The error in case 5 is between 3.8 to 92.2 times smaller than in case 4, depending on the refractive index, when 17256 uniform subvolumes are used ($\Delta/d$ decreases from 5.0 to 0.05 compared to case 4). This confirms that the shape error is dominant when dealing with large sphere diameter to gap ratio $D/d$ in the near-field regime of thermal radiation. An error less than 5% is achieved for $m_a$ to $m_d$ when using up to 39024 uniform subvolumes. With a nonuniform discretization, the errors for $m_e$ and $m_f$ are respectively 12.4% (72264 subvolumes) and 5.0% (52388 subvolumes).

### C. Regime $k_0D \gg 1$

A size of $k_0D = 5.03$ ($D = 8$ μm) and two gap thicknesses of $d/\lambda = 0.01$ ($d = 100$ nm) and $d/\lambda = 0.5$ ($d = 5$ μm) are considered (cases 6 and 7, respectively). Note that case 7 corresponds to the transition between the near- and far-field regimes of thermal radiation. The spatial distribution of the power absorbed for cases 6 and 7 is quite different from the previous problems analyzed. Here, the absorption distribution depends strongly on the imaginary part of the refractive index. The normalized volumetric power absorbed in case 6 for the refractive indices $m_c$ and $m_d$, which have the same real part but have different imaginary parts, are compared in Fig. 10 when 33552 uniform subvolumes per sphere are used. It is important to note that even if the gap $d$ is much smaller than the wavelength $\lambda$ in case 6, propagating modes have a non-negligible contribution to heat transfer since the distance $X(y,z)$ between the spheres varies from 100 nm (near-field) to 8.1 µm (~ far-field). When the imaginary part of the refractive index is large, most of the absorption occurs within the first half of the sphere (Fig. 10(b)). However, the contributing portion of the absorber to the overall heat transfer is larger than for cases 1 to 5 due to the important contribution of propagating modes with larger penetration depth than the evanescent modes. For example, in case 4, where $D/d = 160$ ($D = 1.6$ µm), more than 95% of the power is absorbed within the first 200 nm of the absorber with $m_d$. In case 6, where $D/d = 80$ ($D = 8$ µm), this distance is equal to 4.4 µm with $m_d$. When the imaginary part of the refractive index is small, a different pattern is observed in the power distribution (Fig. 10(a)). For this case, the whole sphere contributes significantly to the overall heat transfer. This is because the thermally generated propagating waves experience multiple reflections within the sphere due to low absorption. Yet, it can be seen in Fig. 10(a) that significant absorption occurs within a small portion of the sphere facing the emitter due to evanescent modes with minimum penetration depth approximately equal to $d$. It is thus clear that the proximity approximation cannot be applied for cases where the imaginary part of the refractive index is small, since the absorber is optically thin. Additionally, nonuniform discretization for cases 6 and 7 is not as effective as for the previous cases, since a large portion of the absorber contributes to heat transfer. A fine discretization is thus required throughout the spheres. For this reason, cases 6 and 7 are difficult to handle with the T-DDA.

The absolute value of the relative error is shown in Fig. 11 for case 6. The errors for $m_a$ and $m_f$ are 2.3% and 2.75%, respectively, when 33552 uniform subvolumes are used. This confirms that when $|m|$ is small, the T-DDA is accurate regardless of the parameters of the problem. The error grows rapidly as $|m|$ increases such that an error of 96.4% is obtained with 82712 uniform subvolumes for $m_e$. Nonuniform discretization has been applied to refractive indices $m_b$ through $m_e$. Since the gap size is small, a fine discretization is required at the front side of the spheres. Yet, a fine discretization also needs to be applied to the whole sphere (when the imaginary part of the refractive index is small) or to a large portion of the absorber (when the imaginary part of the refractive index is large). As shown in Fig. 11, the error reduces considerably when nonuniform discretization is used. Errors less than 5% are obtained for $m_b$ and $m_d$ using 48367 nonuniform subvolumes. Furthermore, the error for $m_e$ reduces to 17.0% with 81980 nonuniform subvolumes. The smallest error obtained for $m_c$ is 22.1% with 60200 nonuniform subvolumes. The number of subvolumes used for $m_c$ should be increased further if a better accuracy is desired.

The absolute value of the relative error for case 7 is provided in Fig. 12, where the gap size is increased to $d/\lambda = 0.5$ ($d = 5$ µm). As expected, the errors are considerably smaller than in case 6 due to the larger gap size which results in a smaller $\Delta/d$ and thus a smaller shape error. For the refractive indices $m_a$, $m_b$, and $m_f$, an error approximately equal to, or smaller than, 5% is obtained with 2176 uniform subvolumes per sphere. Nonuniform discretization has been applied to the other refractive indices. Since the distance between the spheres $X(y,z)$ vary between 5 µm and 13 µm, most of the energy is transferred by propagating modes such that there is a larger contribution from the back side of the absorber to the overall heat exchange when compared to case 6. For example, for the refractive index $m_d$, approximately 95% of the absorption happens within the first 7.4 µm of the absorber while this distance was equal to 4.4 µm in case 6. Therefore, the nonuniform discretization is not as efficient as in case 6. Errors of 3.8% (66796 nonuniform subvolumes), 8.3% (67472 nonuniform subvolumes), and 27.8% (70544 nonuniform subvolumes) are obtained for $m_c$, $m_d$, and $m_e$, respectively.

## V. CONCLUSIONS

The accuracy and convergence of the T-DDA was analyzed using the exact solution for two spheres separated by a vacuum gap. The study was performed as a function of the size, the gap size and the refractive index. The key results of the convergence analysis are summarized in Table III and the main conclusions are:

1. An error less than 5% was obtained for 74% of the cases studied using up to 82712 subvolumes.

2. Nonuniform discretization is particularly useful when the sphere diameter to gap ratio, $D/d$, is large, when $d \ll \lambda$ and $D < \lambda$, such that significant absorption occurs within a small portion of the sphere. Additionally, nonuniform discretization mitigates the shape error by allowing a better representation of the variation of the gap size by decreasing $\Delta/d$ at the front side of the spheres without increasing drastically the number of subvolumes. The value of $\Delta/d$ leading to a convergent solution varies strongly with the refractive index. For the simulations performed in this study, $\Delta/d \approx 1$ can be satisfactory for the smallest refractive index (1.33 + 0.01$i$) while $\Delta/d \approx 0.01$ is needed for the largest refractive index (3 + 3$i$).

3. For all sizes, the accuracy of the T-DDA decreases as both the real and the imaginary parts of the refractive index increase. A large refractive index affects the accuracy of the results by increasing the variation of the electric field and the free space DGF inside the spheres and by amplifying the shape error. It was also shown that fast convergence is achieved when dealing with resonant modes. The T-DDA is therefore accurate for predicting surface phonon-polariton mediated near-field radiative heat transfer.

4. When the sphere diameter $D$ and the gap size $d$ have the same order of magnitude as the wavelength $\lambda$, nonuniform discretization is not as efficient as for the other cases. For this situation, the whole sphere contributes to the overall heat transfer such that a fine discretization is required throughout the entire volume of the absorber.

The conclusions of this paper are applicable to other geometries, except that the error is likely to be smaller due to a weaker shape error. The T-DDA is currently suitable for particles with sizes smaller than, or of the same order of magnitude as, the wavelength due to computational limitations. The accuracy of the T-DDA can potentially be improved further using the various techniques proposed in the DDA literature such as the weighted discretization approach [49,50] and the filtered coupled-dipole method [50-52]. This is left for a future research effort.

## ACKNOWLEDGMENTS

This work was sponsored by the US Army Research Office under Grant no. W911NF-14-1-0210. The authors also acknowledge the Extreme Science and Engineering Discovery Environment (NSF grant no. ACI-1053575) and the Center for High Performance Computing at the University of Utah for providing the computational resources used in this study. Financial support for T.T. and R.B. was provided by the Undergraduate Research Opportunities Program (UROP) at the University of Utah and by the NSF Research Experiences for Undergraduates (REU) program (Grant no. CBET-1253577), respectively.

# APPENDIX A. DERIVATION OF THE DIPOLE APPROXIMATION FROM THE T-DDA

In this Appendix, it is shown that the T-DDA applied to the two-sphere problem described in Section IV reduces to the dipole approximation when $D \ll \lambda$ and $d \gg D$ [55]. The first sphere is assumed to be at a temperature $T > 0$ K (emitter), while the second sphere is maintained at 0 K (absorber). In the dipolar regime, each sphere is modeled by a single subvolume behaving as an electric point dipole. The first subvolume is assigned to the emitting sphere, while the second one is allocated to the absorbing sphere. The quantity of interest is the power absorbed in sphere 2 calculated from Eq. (13) using the correlation matrix of the induced dipole moment. The (total) dipole moment in subvolume 2 is related to the (total) dipole moment in subvolume 1 by applying Eq. (10) to subvolume 2:

$$\mathbf{p}_2 = \frac{\alpha_2 k_0^2}{\varepsilon_0} \overline{\overline{\mathbf{G}}}_{21} \cdot \mathbf{p}_1 \tag{A.1}$$

where $\mathbf{p}_2^{fl} = \mathbf{0}$ since subvolume 2 is non-emitting. Equation (A.1) implies that the dipole moment in subvolume 2 is induced by the dipole moment in subvolume 1. The dipole moment in subvolume 1 is also determined using Eq. (10):

$$\mathbf{p}_1 = \frac{3\alpha_1}{(\varepsilon_1 + 2)\alpha_1^{CM}} \mathbf{p}_1^{fl} + \frac{\alpha_1 k_0^2}{\varepsilon_0} \overline{\overline{\mathbf{G}}}_{12} \cdot \mathbf{p}_2 \tag{A.2}$$

According to Eq. (A.2), the total dipole moment of subvolume 1 is the summation of the contributions from the thermally fluctuating dipole moment and the dipole moment induced by subvolume 2 (multiple scattering). In the dipolar regime, the second contribution is assumed to be negligible compared to the first one [55] such that the dipole moment of subvolume 1 is approximated by:

$$\mathbf{p}_1 \approx \frac{3\alpha_1}{(\varepsilon_1 + 2)\alpha_1^{CM}} \mathbf{p}_1^{fl} \tag{A.3}$$

Substituting Eq. (A.3) into Eq. (A.1) and applying the fluctuation-dissipation theorem, the ensemble average of the correlation matrix of the dipole moment of subvolume 2 is given by:

$$\langle \mathbf{p}_2 \otimes \mathbf{p}_2 \rangle \approx \frac{4k_0^4}{\pi \omega \varepsilon_0^2} |\alpha_2|^2 \operatorname{Im}(\alpha_1^{CM}) \Theta(\omega,T) \overline{\overline{\mathbf{G}}}_{21} \cdot \left(\overline{\overline{\mathbf{G}}}_{21}\right)^\dagger \tag{A.4}$$

where $\mathbf{p}_2 = \mathbf{p}_2^{ind}$. The trace of the correlation matrix of the induced dipole moment in subvolume 2 is obtained using Eq. (A.4) and by substituting the free space DGF:

$$\operatorname{tr}\left(\langle \mathbf{p}_2^{ind} \otimes \mathbf{p}_2^{ind} \rangle\right) \approx \frac{\Theta(\omega,T) \operatorname{Im}(\alpha_1^{CM}) |\alpha_2^{CM}|^2 k_0^6}{2\pi^3 \omega \varepsilon_0^2} \left\{ \frac{3}{\left[k_0(d+D)\right]^6} + \frac{1}{\left[k_0(d+D)\right]^4} + \frac{1}{\left[k_0(d+D)\right]^2} \right\} \tag{A.5}$$

Note that when deriving Eq. (A.5), it is assumed that $\alpha_i \approx \alpha_i^{CM}$ ($i$ = 1, 2). It can be seen from Eqs. (11a) and (11b) that as the size of a subvolume decreases, the radiative polarizability approaches the Clausius–Mossotti polarizability such that they are approximately equal in the dipolar regime. The power absorbed by sphere 2 is finally obtained by substituting Eq. (A.5) into Eq. (13). Note that the term $(2/3) k_0^3$ in Eq. (13) has been ignored in some previous DDA formulations due to its small contribution [38,39]. Following the same procedure, the power absorbed in subvolume 2 is given by:

$$\langle Q_{abs,\omega}\rangle \approx \frac{1}{4\pi^3\varepsilon_0^2}\Theta(\omega,T)\operatorname{Im}(\alpha_1^{CM})\operatorname{Im}(\alpha_2^{CM})k_0^6\left\{\frac{3}{\left[k_0(d+D)\right]^6}+\frac{1}{\left[k_0(d+D)\right]^4}+\frac{1}{\left[k_0(d+D)\right]^2}\right\} \quad (A.6)$$

which is the same as the power absorbed derived by Chapuis et al. [55] in the dipolar regime.

## REFERENCES

[1] B. Guha, C. Otey, C. B. Poitras, S. Fan, and M. Lipson, Nano Lett. **12**, 4546 (2012).

[2] Y. De Wilde, F. Formanek, R. Carminati, B. Gralak, P.-A. Lemoine, K. Joulain, J.-P. Mulet, Y. Chen, and J.-J. Greffet, Nature **444**, 740 (2006).

[3] E. A. Hawes, J. T. Hastings, C. Crofcheck, and M. P. Mengüç, Opt. Lett. **33**, 1383 (2008).

[4] V. L. Y. Loke and M. P. Mengüç, J. Opt. Soc. Am. A **27**, 2293 (2010).

[5] C. R. Otey, W. T. Lau, and S. Fan, Phys. Rev. Lett. **104**, 154301 (2010).

[6] S. Basu and M. Francoeur, Appl. Phys. Lett. **98**, 113106 (2011).

[7] A. Babuty, K. Joulain, P.-O. Chapuis, J.-J. Greffet, and Y. De Wilde, Phys. Rev. Lett. **110**, 146103 (2013).

[8] A. C. Jones and M. B. Raschke, Nano Lett. **12**, 1475 (2012).

[9] B. T. O'Callahan, W. E. Lewis, A. C. Jones, and M. B. Raschke, Phys. Rev. B **89**, 245446 (2014).

[10] R. S. DiMatteo, P. Greiff, S. L. Finberg, K. A. Young-Waithe, H. K. H. Choy, M. M. Masaki, and C. G. Fonstad, Appl. Phys. Lett. **79**, 1894 (2001).

[11] K. Park, S. Basu, W. P. King, and Z. M. Zhang, J. Quant. Spectrosc. Radiat. Transfer **109**, 305 (2008).

[12] M. Francoeur, R. Vaillon, and M. P. Mengüç, IEEE Trans. Energy Convers. **26**, 686 (2011).

[13] A. Kittel, W. Müller-Hirsch, J. Parisi, S.-A. Biehs, D. Reddig, and M. Holthaus, Phys. Rev. Lett. **95**, 224301 (2005).

[14] E. Rousseau, A. Siria, G. Jourdan, S. Volz, F. Comin, J. Chevrier, and J.-J. Greffet, Nat. Photonics **3**, 514 (2009).

[15] S. Shen, A. Narayanaswamy, and G. Chen, Nano Lett. **9**, 2909 (2009).

[16] S. M. Rytov, Y. A. Kravtsov, and V. I. Tatarskii, *Principles of Statistical Radiophysics 3: Elements of Random Fields*, (Springer, New York, 1989).

[17] D. Polder and M. Van Hove, Phys. Rev. B **4**, 3303 (1971).

[18] K. Joulain, J.-P. Mulet, F. Marquier, R. Carminati, and J.-J. Greffet, Surf. Sci. Rep. **57**, 59 (2005).

[19] C. J. Fu and Z. M. Zhang, Int. J. Heat Mass Transfer **49**, 1703 (2006).

[20] J.-P. Mulet, K. Joulain, R. Carminati, and J-J. Greffet, Nanoscale Microscale Thermophys. Eng. **6**, 209 (2002).

[21] M. Francoeur, M. P. Mengüç, and R. Vaillon, J. Quant. Spectrosc. Radiat. Transfer **110**, 2002 (2009).

[22] R. Messina, M. Tschikin, S.-A. Biehs, and P. Ben-Abdallah, Phys. Rev. B **88**, 104307 (2013).

[23] A. Narayanaswamy and G. Chen, Phys. Rev. B **77**, 075125 (2008).

[24] K. Sasihithlu and A. Narayanaswamy, Opt. Express **19**, 772 (2011).

[25] K. Sasihithlu and A. Narayanaswamy, Phys. Rev. B **83**, 161406(R) (2011).

[26] M. Krüger, T. Emig, and M. Kardar, Phys. Rev. Lett. **106**, 210404 (2011).

[27] C. Otey and S. Fan, Phys. Rev. B **84**, 245431 (2011).

[28] B. Liu and S. Shen, Phys. Rev. B **87**, 115403 (2013).

[29] A. J. Rodriguez, O. Ilic, P. Bermel, I. Celanovic, J. D. Joannopoulos, M. Soljacic, and S. G. Johnson, Phys. Rev. Lett. **107**, 114302 (2011).

[30] A. Datas, D. Hirashima, and K. Hanamura, J. Therm. Sci. Technol. **8**, 91 (2013).

[31] A. Didari and M. P. Mengüç, J. Quant. Spectrosc. Radiat. Transfer **146**, 214 (2014).

[32] S.-B. Wen, J. Heat Transfer **132**, 072704 (2010).

[33] A. W. Rodriguez, M. T. H. Reid, and S. G. Johnson, Phys. Rev. B **86**, 220302(R) (2012).

[34] A. P. McCauley, M. T. H. Reid, M. Krüger, and S. G. Johnson, Phys. Rev. B **85**, 165104 (2012).

[35] S. Edalatpour and M. Francoeur, J. Quant. Spectrosc. Radiat. Transfer. **133**, 364 (2014).

[36] E. M. Purcell and C. R. Pennypacker, Astrophys. J. **186**, 705 (1973).

[37] B. T. Draine and P. J. Flatau, J. Opt. Soc. Am. A **11**, 1491 (1994).

[38] M. A. Yurkin and A. G. Hoekstra, J. Quant. Spectrosc. Radiat. Transfer **106**, 558 (2007).

[39] B. T. Draine, Astrophys. J. **333**, 848 (1988).

[40] I. Ayrancı, R. Vaillon, and N. Selçuk, J. Quant. Spectrosc. Radiat. Transfer **103**, 83 (2007).

[41] M. A. Yurkin, D. De Kanter, and A. G. Hoekstra, J. Nanophotonics **4**, 041585 (2010).

[42] P. J. Flatau, K. A. Fuller, and D. W. Mackowski, Appl. Opt. **32**, 3302 (1993).

[43] Y.-L. Xu and B. A. S. Gustafson, Astrophys. J. **513**, 894 (1999).

[44] M. A. Yurkin, V. P. Maltsev, and A. G. Hoekstra, J. Opt. Soc. Am. A **23**, 2578 (2006).

[45] M. A. Yurkin, V. P. Maltsev, and A. G. Hoekstra, J. Opt. Soc. Am. A **23**, 2592 (2006).

[46] A. C. Andersen, H. Mutschke, T. Posch, M. Min, and A. Tamanai, J. Quant. Spectrosc. Radiat. Transfer **100**, 4 (2006).

[47] B. T. Draine and P. J. Flatau, arXiv:1305.6497.

[48] M. A. Yurkin and A. G. Hoekstra, J. Quant. Spectrosc. Radiat. Transfer **112**, 2234 (2011).

[49] N. B. Piller, Opt. Lett. **22**, 1674 (1997).

[50] N. B. Piller, Opt. Comm. **160**, 10 (1999).

[51] N. B. Piller and O. J. F. Martin, IEEE Trans. Ant. Propag. **46**, 1126 (1998).

[52] M. A. Yurkin, M. Min, and A. G. Hoekstra, Phys. Rev. E **82**, 036703 (2010).

[53] M. A. Yurkin and M. Kahnert, J. Quant. Spectrosc. Radiat. Transfer **123**, 176 (2013).

[54] E. Zubko, D. Petrov, Y. Grynko, Y. Shkuratov, H. Okamoto, K. Muinonen, T. Nousiainen, H. Kimura, T. Yamamoto, and G. Videen, Appl. Opt. **49**, 1267 (2010).

[55] P.-O. Chapuis, M. Laroche, S. Volz, and J.-J. Greffet, Appl. Phys. Lett. **92**, 201906 (2008).

[56] L. Novotny and B. Hecht, *Principles of Nano-Optics* (Cambridge University Press, New York, 2006).

[57] W. C. Chew, *Waves and Fields in Inhomogeneous Media* (IEEE Press, Piscataway, 1995).

[58] P. C. Chaumet, A. Sentenac, and A. Rahmani, Phys. Rev. E **70**, 036606 (2004).

[59] D. A. Smunev, P. C. Chaumet, and M. A. Yurkin, J. Quant. Spectrosc. Radiat. Transfer **156**, 67 (2015).

[60] D. J. Griffiths, *Introduction to Electrodynamics* (Prentice Hall, Upper Saddle River, 1999).

[61] R. Schmehl, MS Thesis, Arizona State University, 1994.

[62] R. D. Yates and D. J. Goodman, *Probability and Stochastic Processes* (Wiley, Hoboken, 2005), 2$^{nd}$ edition.

[63] A. Lakhtakia and G. W. Mulholland, J. Res. Nat. Inst. Stand. Technol. **98**, 699 (1993).

[64] C. E. Dungey and C. F. Bohren, J. Opt. Soc. Am. A **8**, 81 (1991).

[65] B. T. Draine and J. Goodman, Astrophys. J. **405**, 685 (1993).

[66] D. Gutkowicz-Krusin and B. T. Draine, arXiv:astro-ph/0403082.

[67] A. Rahmani, P. C. Chaumet, and G. W. Bryant, Opt. Lett. **27**, 2118 (2002).

[68] A. Rahmani, P. C. Chaumet, and G. W. Bryant, Astrophys. J. **607**, 873 (2004).

[69] M. J. Collinge and B. T. Draine, J. Opt. Soc. Am. A **21**, 2023 (2004).

[70] P. C. Chaumet, A. Rahmani, F. de Fornel, and J.-P. Dufour, Phys. Rev. B **58**, 2310 (1998).

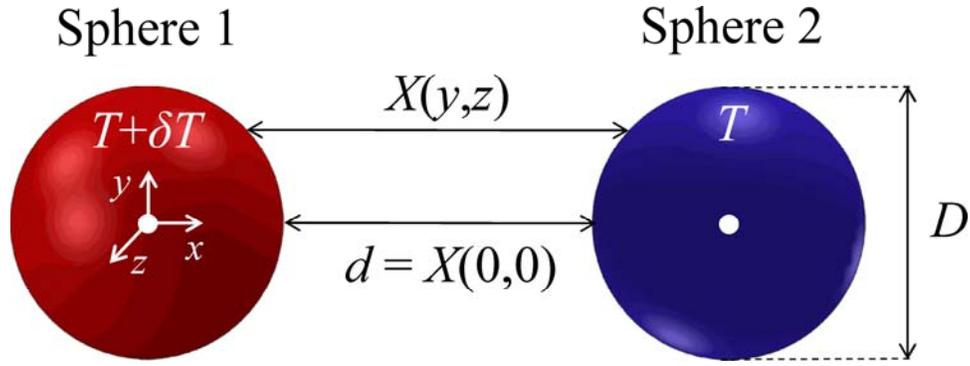

FIG. 1. (Color online) Schematic of the problem under consideration: two spheres of diameter $D$ separated by a distance $X(y,z)$ are exchanging thermal radiation. The minimum distance between the spheres is $d$ (= $X(0,0)$).

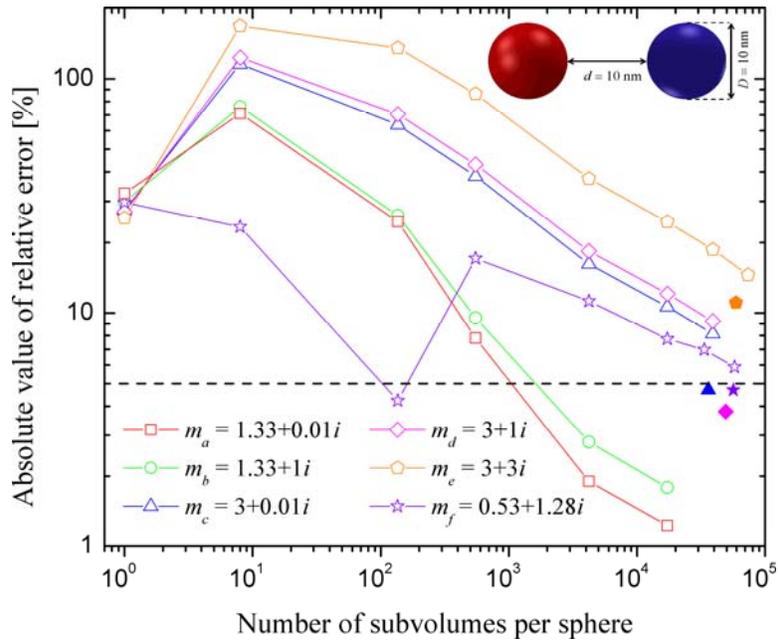

FIG. 2. (Color online) Absolute value of the relative error of the conductance as a function of the number of subvolumes and the refractive index for case 1 ($k_0D = 0.00628$, $d/\lambda = 0.001$). Open and filled symbols denote the error for uniform and nonuniform discretizations, respectively.

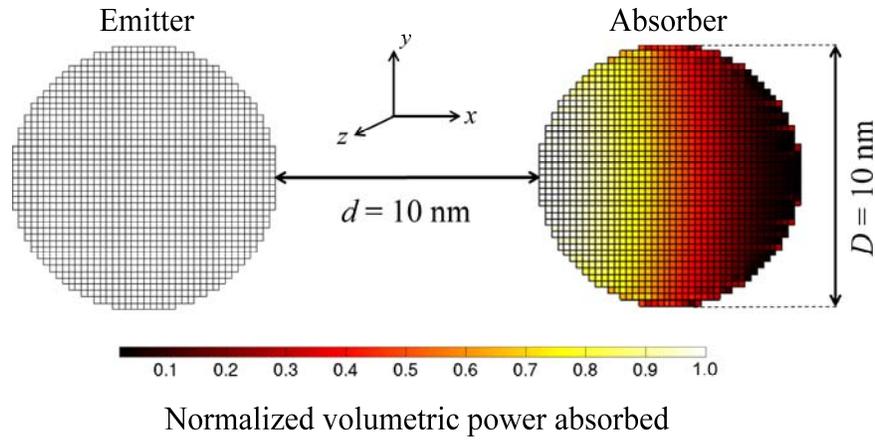

Normalized volumetric power absorbed

FIG. 3. (Color online) Spatial distribution of the normalized volumetric power absorbed for case 1 (refractive index $m_c$, 39024 uniform subvolumes).

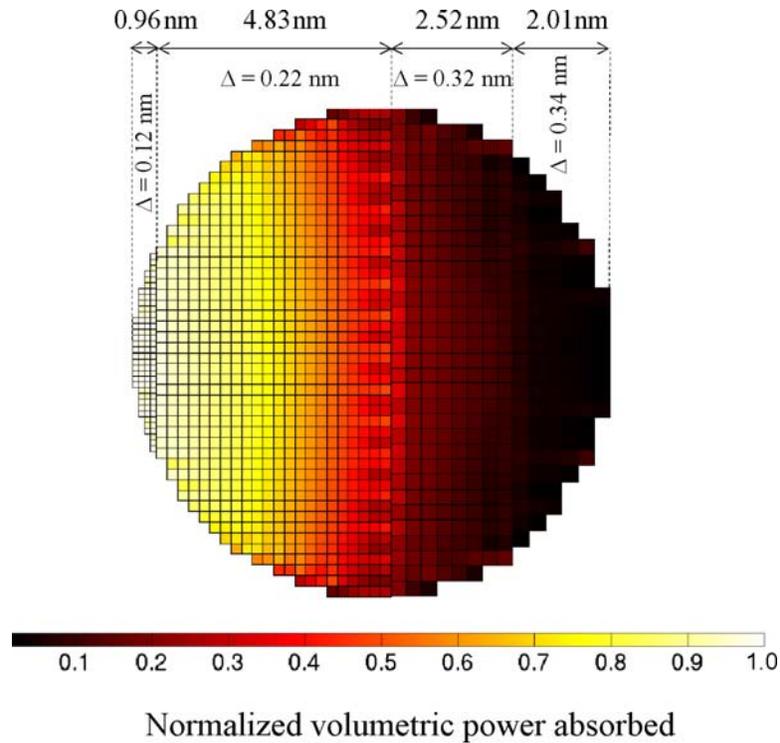

Normalized volumetric power absorbed

FIG. 4. (Color online) Spatial distribution of the normalized volumetric power absorbed for case 1 (refractive index $m_c$, 36168 nonuniform subvolumes); the size of the subvolumes increases as the power absorbed decreases.

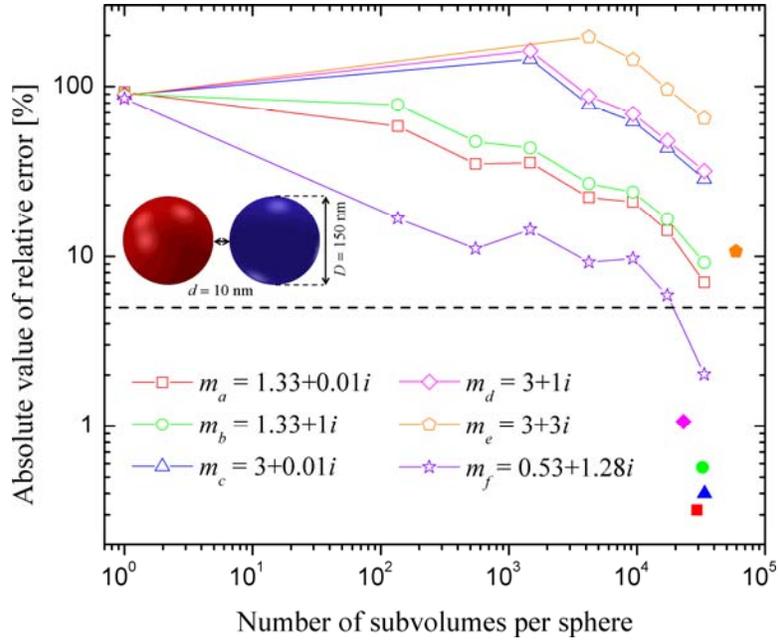

FIG. 5. (Color online) Absolute value of the relative error of the conductance as a function of the number of subvolumes and the refractive index for case 2 ($k_0D = 0.0943$, $d/\lambda = 0.001$). Open and filled symbols denote the error for uniform and nonuniform discretizations, respectively.

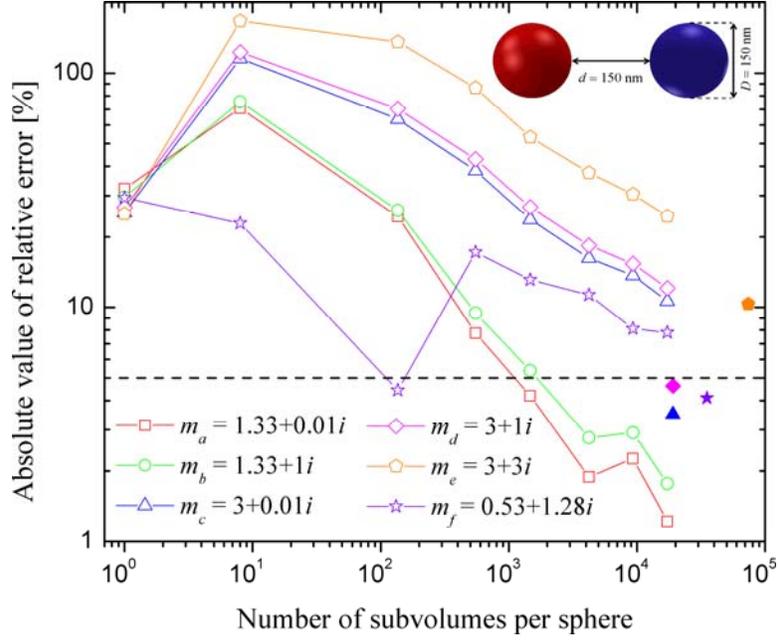

FIG. 6. (Color online) Absolute value of the relative error of the conductance as a function of the number of subvolumes and the refractive index for case 3 ($k_0 D = 0.0943$, $d/\lambda = 0.015$). Open and filled symbols denote the error for uniform and nonuniform discretizations, respectively.

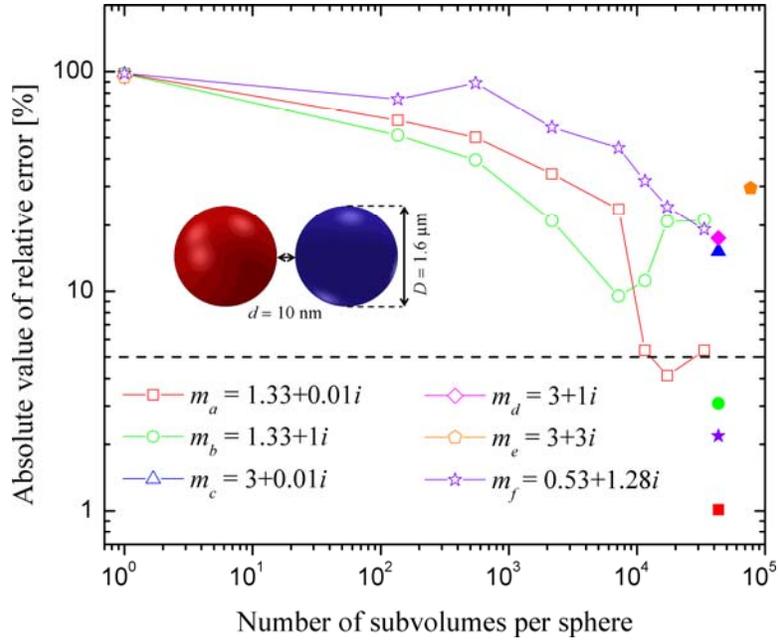

FIG. 7. (Color online) Absolute value of the relative error of the conductance as a function of the number of subvolumes and the refractive index for case 4 ($k_0 D = 1.01$, $d/\lambda = 0.001$). Open and filled symbols denote the error for uniform and nonuniform discretizations, respectively.

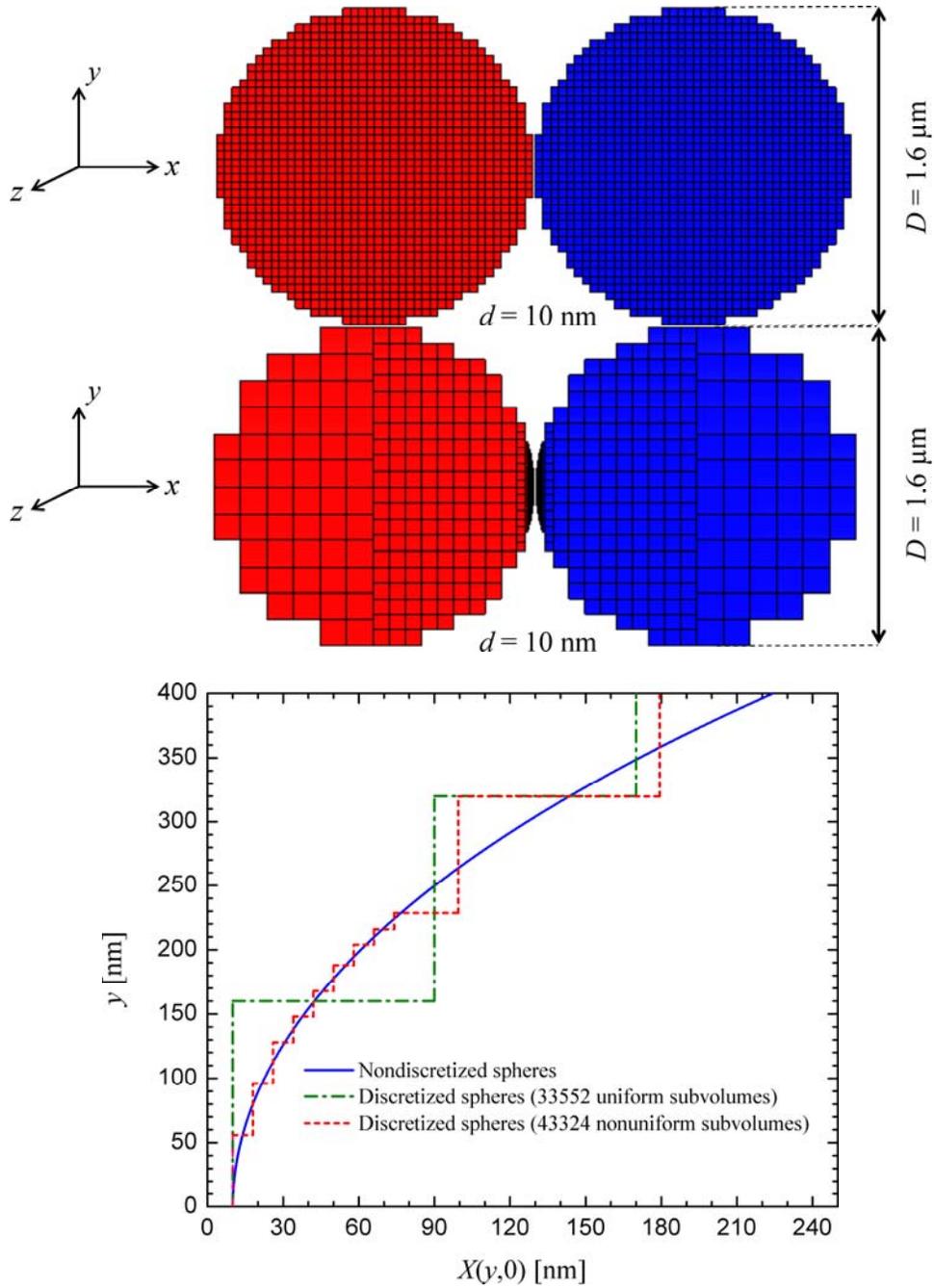

FIG. 8. (Color online) Two spheres of diameter $D = 1.6$ μm separated by a gap size $d = 10$ nm are discretized using (a) 33552 uniform subvolumes, and (b) 43324 nonuniform subvolumes. (c) Variation of the distance between the spheres, $X(y,0)$, along the $y$-axis for the uniform and nonuniform discretizations shown in panels (a) and (b).

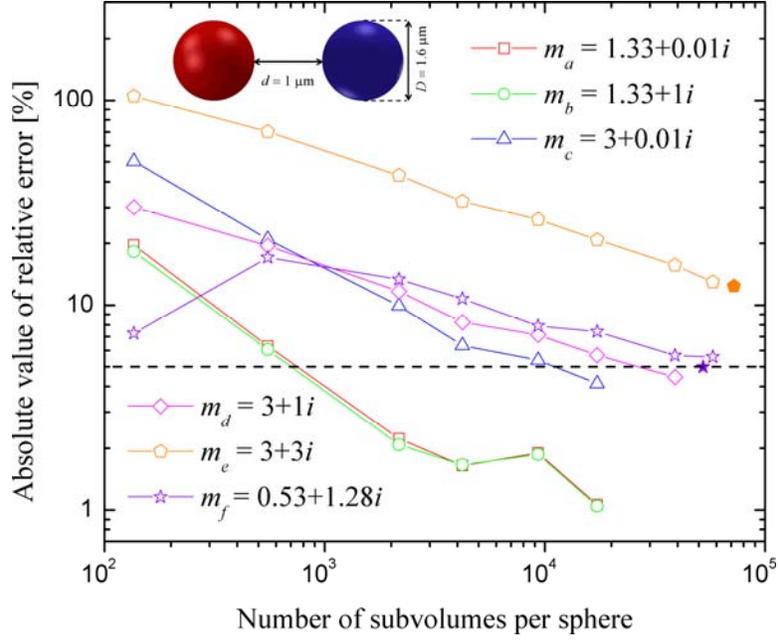

FIG. 9. (Color online) Absolute value of the relative error of the conductance as a function of the number of subvolumes and the refractive index for case 5 ($k_0 D = 1.01$, $d/\lambda = 0.1$). Open and filled symbols denote the error for uniform and nonuniform discretizations, respectively.

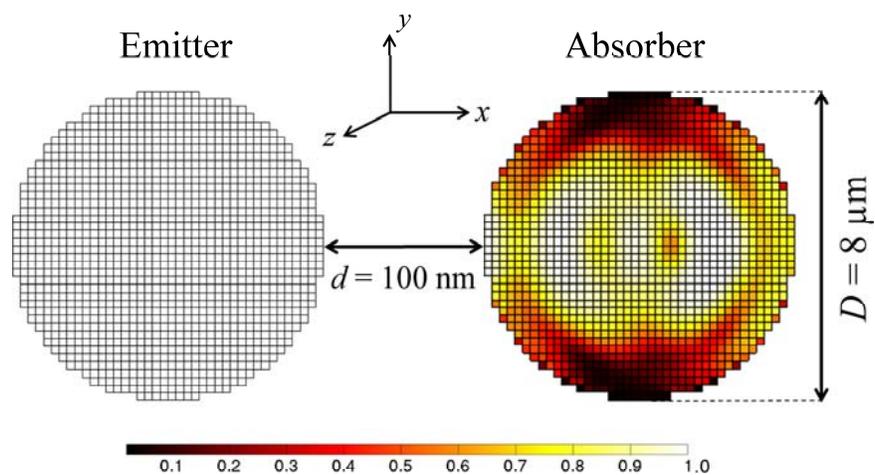

Normalized volumetric power absorbed

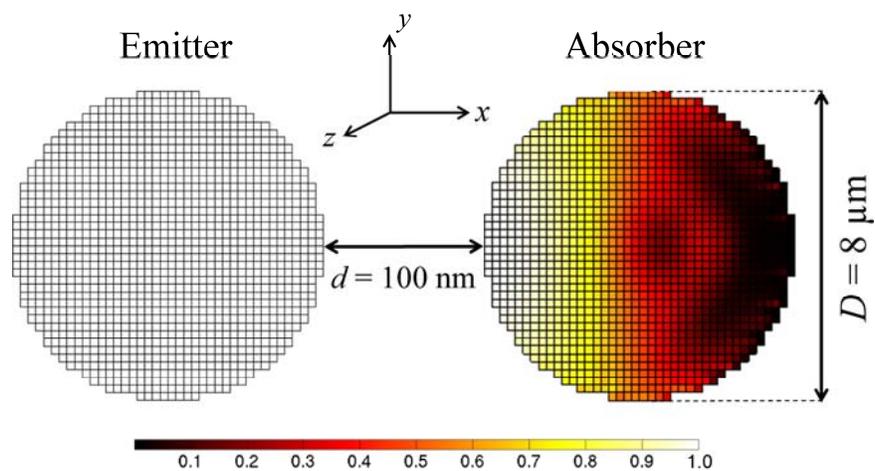

Normalized volumetric power absorbed

FIG. 10. (Color online) Spatial distribution of the normalized volumetric power absorbed for case 6 (33552 uniform subvolumes): (a) Refractive index $m_c$ = 3+0.01$i$, and (b) refractive index $m_d$ = 3+1$i$.

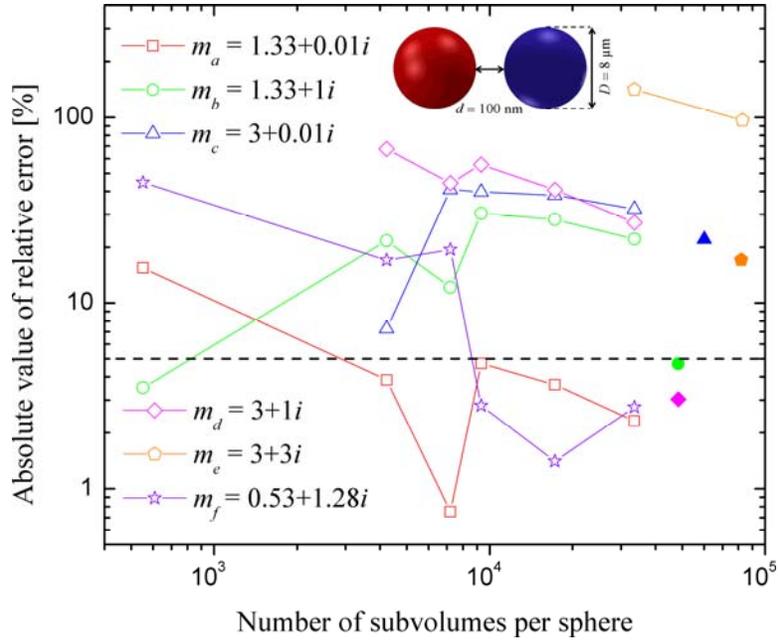

FIG. 11. (Color online) Absolute value of the relative error of the conductance as a function of the number of subvolumes and the refractive index for case 6 ($k_0 D = 5.03$, $d/\lambda = 0.01$). Open and filled symbols denote the error for uniform and nonuniform discretizations, respectively.

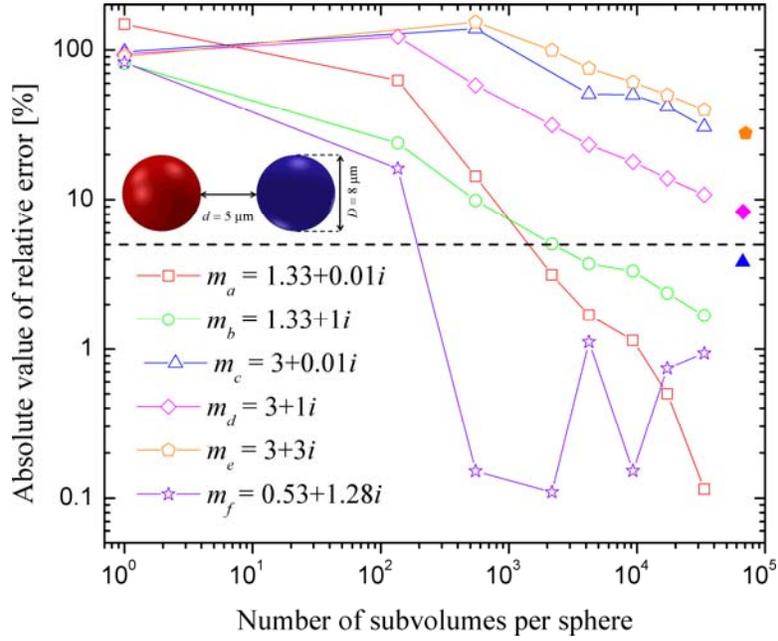

FIG. 12. (Color online) Absolute value of the relative error of the conductance as a function of the number of subvolumes and the refractive index for case 7 ($k_0 D = 5.03$, $d/\lambda = 0.5$). Open and filled symbols denote the error for uniform and nonuniform discretizations, respectively.

TABLE I. Cases investigated in the convergence analysis.

|  | $k_0 D$ ($D$) | $d/\lambda$ ($d$) |
|---|---|---|
| Case 1 | 0.00628 (10 nm) | 0.00100 (10 nm) |
| Case 2 | 0.0943 (150 nm) | 0.00100 (10 nm) |
| Case 3 | 0.0943 (150 nm) | 0.0150 (150 nm) |
| Case 4 | 1.01 (1.6 μm) | 0.00100 (10 nm) |
| Case 5 | 1.01 (1.6 μm) | 0.100 (1 μm) |
| Case 6 | 5.03 (8 μm) | 0.0100 (100 nm) |
| Case 7 | 5.03 (8 μm) | 0.500 (5 μm) |

TABLE II. Refractive indices investigated in the convergence analysis.

|  |  |
|---|---|
| $m_a$ ($\varepsilon_a$) | 1.33+0.01$i$ (1.77+0.0266$i$) |
| $m_b$ ($\varepsilon_b$) | 1.33+1$i$ (0.769+2.66$i$) |
| $m_c$ ($\varepsilon_c$) | 3+0.01$i$ (9+0.06$i$) |
| $m_d$ ($\varepsilon_d$) | 3+1$i$ (8+6$i$) |
| $m_e$ ($\varepsilon_e$) | 3+3$i$ (0+18$i$) |
| $m_f$ ($\varepsilon_f$) | 0.53+1.28$i$ (-1.36+1.36$i$) |

TABLE III. Smallest relative error of the conductance obtained for all cases considered in the convergence analysis ($N$: number of subvolumes per sphere; U: uniform discretization; NU: nonuniform discretization).

|  | Case 1 | | Case 2 | | Case 3 | | Case 4 | | Case 5 | | Case 6 | | Case 7 | |
|---|---|---|---|---|---|---|---|---|---|---|---|---|---|---|
|  | $N$ | Error (%) | $N$ | Error (%) | $N$ | Error (%) | $N$ | Error (%) | $N$ | Error (%) | $N$ | Error (%) | $N$ | Error (%) |
| $m_a$ | 17256 (U) | 1.23 | 29340 (NU) | 0.318 | 17256 (U) | 1.22 | 43324 (NU) | 1.01 | 17256 (U) | 1.06 | 33552 (U) | 2.31 | 33552 (U) | 0.115 |
| $m_b$ | 17256 (U) | 1.79 | 32572 (NU) | 0.572 | 17256 (U) | 1.77 | 43324 (NU) | 3.08 | 17256 (U) | 1.04 | 48368 (NU) | 4.70 | 33552 (U) | 1.68 |
| $m_c$ | 36168 (NU) | 4.70 | 33740 (NU) | 0.405 | 19064 (NU) | 3.50 | 43324 (NU) | 15.2 | 17256 (U) | 4.14 | 60200 (NU) | 22.1 | 66796 (NU) | 3.83 |
| $m_d$ | 49216 (NU) | 3.80 | 23080 (NU) | 1.06 | 19064 (NU) | 4.62 | 43324 (NU) | 17.5 | 39024 (U) | 4.45 | 48368 (NU) | 3.03 | 67472 (NU) | 8.33 |
| $m_e$ | 59360 (NU) | 11.1 | 59408 (NU) | 10.7 | 74180 (NU) | 10.3 | 77196 (NU) | 29.5 | 72264 (NU) | 12.4 | 81980 (NU) | 17.0 | 70544 (NU) | 27.8 |
| $m_f$ | 56500 (NU) | 4.69 | 33552 (U) | 2.01 | 35256 (NU) | 4.10 | 43324 (NU) | 2.18 | 52388 (NU) | 4.99 | 33552 (U) | 2.75 | 33552 (U) | 0.931 |